**The Lindbladian Form**

**And**

**The Reincarnation of Felix Bloch's Generalized Theory of Relaxation**

Thomas M. Barbara

Advanced Imaging Research Center

Oregon Health and Sciences University

Portland, OR  97239

Nov 10th, 2020

Email: barbarat@ohsu.edu

Abstract

The relationship between the classic magnetic resonance density matrix relaxation theories of Bloch and Hubbard, and the modern Lindbladian master equation methods are explored.   These classic theories are in full agreement with the latest results obtained by the modern methods.  A careful scrutiny shows that this also holds true for Redfield's later treatment, offered in 1965.  The early contributions of Bloch and Hubbard to rotating frame relaxation theory are also highlighted.  Taken together, these seminal efforts of Bloch and Hubbard can enjoy a new birth of contemporary relevance in magnetic resonance.

## 1. Introduction

In a recent and important publication [1], Bengs and Levitt formalize relaxation theory for systems that deviate significantly from the equilibrium state, conditions that invalidate the high temperature, weak ordering approximation, a corner stone in the commonly used Redfield theory [2]. Bengs and Levitt employed very modern methods that arose in the late 1970's and are now fundamental in the topic of open quantum systems. These methods are often referred to eponymously as the "Lindbladian Form". In addition to [1] a useful tutorial on this topic can be found in the work of Manzano [3]. Even though many, if not all of the classic papers are cited in their offering of their own Lindbladian analysis, the authors of [1], were not aware that a result identical to theirs was already offered by Bloch in 1957 [4], and masterly expounded upon by Hubbard in 1961 [5]. This conclusion can be ascertained by a glance at Table 1 of [1], where references to neither Bloch, nor Hubbard appear in the rightmost column. From a study of past efforts, as reviewed by Bengs and Levitt along with their own results, this holds for Abragam, for Jeener, and for Ernst [1]. The significance of Bloch and Hubbard has gone unappreciated by the NMR community for decades. By approaching the problem from a new perspective, and importantly, with new experimental measurements on specially prepared spin systems, Bengs and Levitt have resolved a long standing and long misunderstood issue.

This confusion over the results of Bloch and Hubbard is likely due, in part, to the use of difficult notation by Bloch. Hubbard's treatment is a significant improvement but also possesses a few obscure aspects. In that situation, it is not difficult to understand that nearly all NMR researchers rely on the simpler Redfield formalism, especially given the fact that the conditions under which the approximations are applicable, are those encountered in the vast majority of cases. Bengs and Levitt make a detailed comparison between other proposals for a more proper relaxation theory that naturally contains the correct equilibrium steady state [1], and all are found to be defective in one way or another. One argument for the importance of Bengs and Levitt's effort, is their unequivocal and independent confirmation of the success of Bloch and Hubbard over the other formulations. The claim, that a significant aspect of relaxation theory, with an overcast coastline, now enjoys blue sky, is a fair one.

A *brief* exposition of Bloch's main results, by way of Hubbard, using Hubbard's own notation, appears therefore to be a worthwhile endeavor, with historical importance and reasonable expectations that such an effort will be of some interest to NMR researchers in general. How this task has fallen into the hands of the author may also be of some help in orienting the reader.

The origins reside in the author's thesis work, a part of which involved the problem of anisotropic spin lattice relaxation measurements for deuterated molecules dissolved in a liquid crystalline matrix. This induced the author to study all the early papers on relaxation theory very carefully. Discovering the significance of Bloch, and the clarity of Hubbard's exposition, left an indelible impression, even though the author happily used the simpler Redfield formalism for the task at hand. The author abandoned the field in 1988 to pursue other interests, but always enjoyed reading the current NMR relaxation literature. Twenty five years later, a coworker gently persuaded the author to work out aspects of relaxation and exchange during adiabatic sweeps. After completing that task, the author reacquainted himself with Bloch and Hubbard in order to understand some lingering details. This effort fully prepared the author to recognize how the work of Bloch and Hubbard tied into, and formed a precedent for, the recent efforts of Bengs and Levitt.

The main goal is then to *go beyond a citation*, and demonstrate the equivalence of the major result in [1] to that of Bloch and Hubbard. The significant and original result offered here, is in the mathematical analysis required to reveal the equivalent success of Bloch and Hubbard towards the problem confronted and solved by Bengs and Levitt by their use of Lindbladian methods. The effort is not trivial, and requires a careful study of Hubbard's notation and the development of symmetry properties that Hubbard does not provide. Furthermore, Hubbard offers expressions that in some ways are superior to the "fully reduced" Lindbladian form, as will be discussed in section 3. In the conclusion section, the relation of these aspects of relaxation theory to later work by Redfield is commented on.

A short discussion of the Lindbladian form is presented beforehand, from a simple and very direct approach which possesses useful didactics, thereby allowing non experts to appreciate the essence of the formalism without going into a full review of the mathematical details involved in a rigorous proof. Since the Lindblad approach is rather new to the NMR literature, it can be helpful to have a more facile presentation. Those readers interested in more details can find valuable sources in Bengs and Levitt, Manzano and Gyamfi [6].

In addition to the Lindbladian form of Bloch's generalized theory of relaxation, we take advantage of the opportunity to highlight Bloch's and Hubbard's early, and also largely unrecognized, contributions to the dynamics of spin locking and rotating frame relaxation. All together these aspects form a basis for a renewed interest, or at the very least, a new and greater appreciation of these classic publications.

**2: A Guide to Lindblad**

The original publications on the Lindblad form are of a very mathematical flavor, both for the case of finite dimensions, discussed by Gorini et al. [7], as well as for a general Hilbert space by Lindblad [8]. Because of these combined efforts, the Lindblad form is also often referred to as the GKSL equation. For finite dimensional problems, as is pertinent to NMR applications, it is possible to offer a straightforward method of construction, that is reasonably motivated and which uses elementary matrix algebra. What follows below is very much ex post facto, and was developed by the author after reading Bengs and Levitt. It is not without precedent however. An interesting historical overview of the Lindbladian form can be found on the physics archives [9]. Therein, one can find examples of nearly identical approaches and even a very early use of the Lindbladian form by Landau in 1927. For a very readable tutorial and overview, [3] and [6] are recommended. For strictly mathematical proofs on the semi-group, [7] and [8] should be consulted. The reader should not interpret what follows as a replacement to the rigorous mathematical if and only if proofs of the semi-group. The key to the approach is in the importance of matrix factoring. Indeed, matrix factoring in terms of Kronecker products is an essential ingredient in the mathematical proofs, but its role is not often explicitly highlighted in the manner given here. Matrix products also arise in a straight forward manner when relaxation theory is approached by way of weak coupling perturbation theory for the spin system and the bath, so that in this case the spin matrices are already factored. Along the way, the simplest Lindbladian, the one used for exchange in NMR, and which involves the factoring of the identity matrix, is presented.

As an *ansatz*, one can start from a *generalized* "state vector" dynamics, governed by a general complex matrix $M$

$$\dot{c}_t = \sum_j M_{ij}\, c_j \qquad (1)$$

In Eq.(1), M generalizes what is usually the Hamiltonian, so that the equation, in a purely analogous manner, represents a type of "time dependent Schrodinger equation" for a finite number of states. From the vector components $c_i$, we construct a Hermitian, rank-one "density matrix" with elements

$$\varrho_{ij} = c_i c_j^* \qquad (2)$$

The square of all rank-one matrices are proportional to themselves, $\varrho^2 = \alpha\varrho$, so that in this case $\alpha = \Sigma c_i c_i^*$. However, keep in mind that the converse does not hold true. The terminology of "rank" has a number of variants, and in this section we adopt the usage common in the theory of matrices as discussed in Halmos [10].

The terms "state vector" and "density matrix" have been placed in quotation marks to emphasize their heuristic labels at this early stage of the construction. The dynamics of this "density matrix" are given by

$$\dot{\varrho} = M\varrho + \varrho M^\dagger \qquad (3)$$

The solution to Eq.(3) can be expressed as

$$\varrho(t) = \Upsilon(t)\varrho(0)\Upsilon(t)^\dagger; \quad \dot{\Upsilon}(t) = M\Upsilon(t); \quad \Upsilon(0) = 1 \qquad (4)$$

Of course, $\Upsilon(t)$ is not unitary since M is not necessarily skew Hermitian. After some type of ensemble averaging, the "density matrix" will *no longer be rank one* and will take on the character of a general Hermitian matrix. This construction is a simple adaptation of the procedure delineated by Landau and Lifschitz [11]. Conservation of probability $Tr(\dot{\varrho}) = 0$ does not hold for Eq.(3), but we are now just two steps away from rectifying that shortcoming. An application of *Cartesian decomposition* allows any matrix to be written in the form

$$M = -\tfrac{1}{2}H + A \qquad (5)$$

H is Hermitian, $H = H^\dagger$ and $A$ is anti-Hermitian, $A = -A^\dagger$, which explains why the notation has been used. The reason for the particular numerical factor of -1/2 in front of H will be revealed below. The dynamics can then be expressed in terms of commutators [x,y] and anti-commutators {x,y}

$$\dot{\varrho} = -\tfrac{1}{2}\{H, \varrho\} + [A, \varrho] \qquad (6)$$

Only the anti-commutator term contributes to $Tr(\dot{\varrho}) = -Tr(H\varrho)$. If $H$ possesses a non-trivial factoring such that $H = NN^\dagger$, conservation of probability can be non-trivially restored by adding a term in $N^\dagger \varrho N$, due to the cyclic properties of the trace operation $Tr(N^\dagger \varrho N) = Tr(NN^\dagger \varrho)$. The augmented dynamics are then transformed into the famous Lindbladian form:

$$\dot{\varrho} = -\tfrac{1}{2}\{NN^\dagger, \varrho\} + [A, \varrho] + N^\dagger \varrho N \qquad (7)$$

The matrix factoring adopted above is known in the numerical matrix analysis literature as Cholesky decomposition [12] and is closely related to polar decomposition, where an arbitrary linear transformation can be written as a product of a positive matrix and an isometry [10]. While positive, it may not be completely positive. The condition of complete positivity requires that the Kronecker product of the matrix with the identity matrix of arbitrary dimension also be positive [3]. This condition has been essential for a strict mathematical proof and also ensures that the density matrix has real, positive eigenvalues. Complete positivity has been critiqued in [13] and [14].

According to Eq.[4], the commutator part *and* the anti-commutator part of Eq.[7] can be eliminated from Eq.[7] by the usual method of performing an interaction contact transformation, which then produces a new dynamical equation $\dot\varrho' = kV^\dagger(t)\varrho'V(t)$. The choice of scaling in Eq.[6] is made so that $k$=1. Throughout the steps used to construct Eq.[7], no assumptions regarding the time dependence of the operators are made, and therefore a semi-group structure, where the total propagation must satisfy $T(t_2 + t_1) = T(t_2)T(t_1)$ does not constitute an essential requirement in this *construction* of the Lindbladian form. This provides some evidence that the exercise offered above is something more than merely an effort to cut corners.

It is interesting to compare conservation of probability for the density matrix, $Tr\dot\varrho = 0$, with that for pure states, $\frac{d}{dt}\sum c_i c_i^* = 0$. This is a more strict condition and holds if and only if $M^\dagger = -M$ as is usual for unitary quantum state evolution with conservation of probability where $\sum c_i c_i^* = 1$ and M can be written as $i\mathcal{H}$ where $\mathcal{H}$ is now the Hamiltonian. By incorporating this condition into a higher dimensional structure, we enjoy greater latitude in having Hermiticity and probability conservation along with richer dynamics that can represent relaxation effects.

The anti-Hermitian part of Eq.(5) has a trace preserving, Hamiltonian evolution and can be taken to represent relaxation induced, or so called, *dynamic frequency shifts*. These are often small in NMR applications, but not exclusively so. A review of these effects in NMR can be found in [15], and this aspect will not be pursued further here.

The Lindbladian form is often written as above, but it is important to recognize that it can be expressed in terms of commutators, as was used in the original work by Gorini et.al.

$$[N, \varrho N^\dagger] + [N\varrho, N^\dagger] = -\{N^\dagger N, \varrho\} + 2N\varrho N^\dagger \qquad (8)$$

In this equation, the alternative factoring, $H = N^\dagger N$ has been adopted. If N is a normal matrix, $[N^\dagger, N] = 0$, the ordering is not important, and in this case, the identity matrix, which corresponds to an infinite temperature limit to within a scalar factor, is a steady state. A similar example is afforded by the factoring given by $H = \{H_1, H_2\} = H_1 H_2 + H_2 H_1$ where both matrices are Hermitian. The resulting expression can be written in terms of nested commutators:

$$[[H_1, \varrho], H_2] + [[H_2, \varrho], H_1] \qquad (9)$$

The anti-commutator can also be re-expressed as a difference between $S^2 = (H_1 + H_2)^2$ and $D^2 = (H_1 - H_2)^2$ and the above equation can written as the combination

$$[S\varrho, S^\dagger] + [S, \varrho S^\dagger] - \left([D\varrho, D^\dagger] + [D, \varrho D^\dagger]\right) \qquad (10)$$

Even though *S* and *D* are both Hermitian, Hermitian conjugates have been kept explicit in order to conform with Eq.[8]. One simple method for constructing a non-normal factoring is to insert the identity between factors. Write $H = P^2 = Pe^{iQ}e^{-iQ}P$ with $[P, Q] \neq 0$ so that $N = Pe^{iQ}$ and $N^\dagger = e^{-iQ}P$. Of course, Q must be Hermitian in order to preserve the Hermitian character of the density matrix in Eq.(7).

A non-normal expansion is generated quite naturally when spectral decomposition is employed. That procedure will be treated in the next section.

While a non-trivial factoring excludes the identity matrix as a sole factor, the identity matrix itself can often be factored. If $H = 1 = RR$ the Lindbladian is now the traditional expression used for the description of intramolecular exchange processes in NMR [16]

$$k(R\varrho R - \varrho) \qquad (11)$$

where in this equation, the constant k is now used to denote the rate of exchange. The case of *intermolecular* exchange is considerably more complicated, and is generally nonlinear, unless high temperatures and small deviations from equilibrium hold forth [17]. It is worth pointing out here that in the literature on the Lindbladian form the N operators used are often referred to as "jump operators" [3].

The use of Cartesian decomposition in Eq.(5) is not actually necessary. If M can be factored, for example, as the product, $AB^\dagger$, then one can directly write down

$$\dot{\varrho} = AB^\dagger\varrho + \varrho BA^\dagger - A^\dagger\varrho B - B^\dagger\varrho A \qquad (12)$$

Or

$$\dot{\varrho} = -\{[B^\dagger\varrho, A] + [A^\dagger, \varrho B]\} \qquad (13)$$

This non-Hermitian, mixed form is useful when comparing Lindbladian expressions with the early relaxation theories of Bloch and Hubbard. In these, the use of spherical tensor operators can obscure the Hermitian character of certain expressions and also produce expressions that on first blush, do not appear to be strictly Lindbladian.

Given that there are $n^2$-1 independent matrices excluding the identity matrix, we can expect to have a linear combination of Linbladian forms, with coefficients that are not related in a rank one fashion, just as with the density matrix. These coefficients represent generalized transport parameters.

Whereas this guide to Lindblad should be sufficient for practical purposes, the author is aware that some readers will be left unsatisfied or even adamantly critical. Therefore, while it was not the author's original intention, some elaboration is offered in the appendix.

The constraints of the Lindbladian form, though powerful, do not provide a complete theory of irreversibility. Other considerations must be brought to bear on the exact manner in which a complicated, many body theory that is fundamentally reversible, can be reduced to a simpler, but now apparently irreversible one. Assumptions of weak coupling between systems (spins and bath) and loss of long time scale correlations as applied to perturbation theory, are common elements throughout both modern Lindbladian and the Bloch Hubbard approaches. Additionally, the need to invoke the secular approximation is paramount all such approaches, as will be discussed in the next section.

## 3. Bloch-Hubbard Relaxation Theory and the Lindbladian Master Equation

As announced in the introduction, is not the author's intention to give an in depth review of the Bloch-Hubbard theory. In particular, Hubbard's exposition is especially clear on most accounts, and those with sufficient interest can directly consult the original publications. Rather, the intention is to provide reasons and the motivation for others to read or revisit these classic works. It is useful to point out here at the outset that many of the mathematical methods used with modern Lindbladian approaches to Markovian systems [1,2] are exactly those used by Bloch and Hubbard, and other aspects of this fact will be emphasized in the conclusions. Since the success of the Bloch-Hubbard theory has gone unrecognized for so many years, and has now been brought into the limelight by the work of Bengs and Levitt, a demonstration of the equivalence can be considered an original contribution to the topic.

We start with Equation [100] of Hubbard's excellent review article of Bloch's generalized theory:

$$R(\sigma) = \sum_{kls} \text{sech}(\beta \omega_s^l/2) J_{lk}(\omega_s^l)\{[O(-\beta)V_s^l O(\beta) \sigma, V^k] - [\sigma O(\beta)V_s^l O(-\beta), V^k]\} \qquad (14)$$

Here we now adopt the notation used by Hubbard and the reader should keep this change in mind. Hubbard uses σ to denote the spin density matrix. The $V_s^l$ operators act on the spin states, and a fuller description of them will be given shortly, as well as how the frequencies $\omega_s^l$ are determined by the eigenvalues of the spin Hamiltonian E. The operator, $O(\beta) = \exp(\frac{\beta E}{2})$ is related to the equilibrium value of the density matrix. Clearly, $R(\sigma)$ is linear in $\sigma$, and as Hubbard points out, it is easy to see that if the density matrix is at equilibrium, $R(\sigma_{eq}) = 0$. The commutator form of Hubbard's equation above is very suggestive. It is almost Lindbladian, but not quite the same as the canonical form.

The sums in Eq.(14) are over integer steps from –n to +n for each index, with different values of n for (k,l) and s. The indexed operators and frequencies satisfy symmetries for negative and positive values of their indices:

$$(V_s^l)^\dagger = V_{-s}^{-l}; \quad \omega_{-s}^{-l} = -\omega_s^l; \quad V^l = \sum_s V_s^l \qquad (15)$$

In order to manipulate Hubbard's expression, we also need some symmetry properties of the $J_{kl}(\omega)$. These index symmetries follow in a straightforward manner from their definitions, which for completeness, and also adhering to Hubbard's original notation, we list.

$$J_{lk}(\omega) = \frac{1}{2}\int_{-\infty}^{\infty} d\tau \ C_{lk}(\tau)e^{i\omega\tau} \qquad (16)$$

Where

$$C_{kl}(\tau) = \frac{1}{2}(A_{kl}(\tau) + A_{kl}(-\tau)) \qquad (17)$$

And

$$A_{kl}(\tau) = Tr_b(\rho^T U^k(\tau)U^l) = Tr_b(\rho^T U^k U^l(-\tau)) \qquad (18)$$

The U operators are bath operators, the Tr_b is a partial trace over bath degrees of freedom with bath equilibrium density matrix ρ^T, and the time dependence is that given by propagation by the bath Hamiltonian. The index symmetries for the J's are then

$$J_{kl}(\omega) = J_{lk}(-\omega) = J_{-k-l}^*(-\omega) \qquad (19)$$

We can now re-sum Hubbard's expression by way of the substitutions $l \rightarrow -l$, $s \rightarrow -s$ in the first term and $k \rightarrow -k$ in the second and using

$$J_{-lk}(\omega_{-s}^{-l}) = J_{l-k}^*(-\omega_{-s}^{-l}) = J_{l-k}^*(\omega_s^l) \qquad (20)$$

The commutators in Eq.(14) are then transformed into

$$[X^\dagger \sigma, V^k] + [(V^k)^\dagger, \sigma X] \qquad (21)$$

Where

$$X = \text{sech}(\beta \omega_s^l / 2) J_{l-k}(\omega_s^l) O(\beta) V_s^l O(-\beta) \qquad (22)$$

We have transformed Hubbard's expression into Lindbladian form of the "non-Hermitian" type as discussed in section 2. In doing so, the ease in demonstrating R($\sigma_{eq}$) = 0 has been lost, but can be restored by combining it with the alternative choice of substitutions $k \rightarrow -k$ in the first term and $l \rightarrow -l$, $s \rightarrow -s$ in the second term, and averaging the two results.

Before reducing Eq.(22) further, an explication of Hubbard's operators is needed. As is common in NMR, the interaction of spin and lattice degrees of freedom are decomposed into products, and indexed in the same manner as Hubbard employs. Hermiticity is enforced by stipulating that operators with indices of opposite sign are Hermitian conjugate to each other. The standard spherical tensor operators of rank L and projection m, are of this type, and while Hubbard does not explicitly indicate this until examples are offered at the end of his article, his $V^k$ operators are basically spherical tensors, where Hubbard uses k to denote the projection index m, and suppresses the rank index L. Likewise, Hubbard is not very explicit regarding his $V_s^l$ operators. He gives their desired properties, but not much on a *general* method for their construction. The key idea is that of spectral decomposition [10] which produces an operator expansion whose coefficients are the eigenvalues of the operator. For the Hamiltonian E we have

$$E = \sum_i \omega_i E_i \qquad (23)$$

Where $\omega_i$ are the eigenvalues of *E* and the $E_i$ are projection operators with the properties

$$E_i E_j = \delta_{ij} E_j \ ; \quad Tr(E_i E_j) = \delta_{ij} \ ; \quad \sum_i E_i = \mathbf{1} \qquad (24)$$

There are a number of methods for constructing the projectors. Perhaps the most straightforward is to use the unitary matrix, U which brings the matrix E to diagonal form. With U at hand we have

$$E_i = U X^{ii} U^\dagger \qquad (25)$$

The fundamental basis matrices have elements $(X^{ij})_{\alpha\beta} = \delta_{i\alpha} \delta_{j\beta}$. A family of matrices can now be constructed from a starting matrix V which will be "eigen-matrices" of the Hamiltonian propagator:

$$[E, E_i V E_j] = (\omega_i - \omega_j) E_i V E_j \qquad (26)$$

$$e^{iEt} E_i V E_j e^{-iEt} = e^{i(\omega_i - \omega_j)t} E_i V E_j \qquad (27)$$

When the V matrices are defined in the eigen-basis of E, this result is almost trivial, for in that case

$$(e^{iEt} V e^{-iEt})_{ij} = e^{i(\omega_i - \omega_j)t} V_{ij} \qquad (28)$$

Even so, the use of projectors allows one to avoid writing out explicit matrix elements. Lexigraphical ordering of these (i,j) index pairs can be adopted and assigned to indices that range from negative to positive integers or odd half integers to obtain Hubbard's $V^l_s$ and his frequencies $\omega^l_s$. The original mathematical lemmas and theorems of Gorini et.al heavily rely on the use of spectral decomposition. Redfield's notation adopts the use of explicit matrix elements, and this perhaps is another of the reasons for the popularity of his equations. Bengs and Levitt commence their own analysis by adopting an eigenbasis for E as with Eq.(28). We can now also tie spectral decomposition to the factoring problem of the previous section. If H is a positive matrix we can apply Eq.(25) to decompose H into a sum

$$H = \sum_k N_k N_k^\dagger \qquad (29)$$

where the operators can be written in terms of the unitary matrix T which diagonalizes H with positive eigenvalues $\lambda_k$ as

$$N_k = \sqrt{\lambda_k} \, T \, X^{kk} \qquad (30)$$

Returning to Hubbard's relaxation expression, one can use the properties of the $V^l_s$ to evaluate the effects of the operator O(β) on $V^l_s$ by employing Eq. [27] with β replacing *it*. We can also expand $V^k$ in terms of $V^l_s$. Finally, we invoke the secular approximation, where rapidly oscillating terms, generated by the evolution of E, are dropped. There are two paths that one can follow for this goal. One path takes the Zeeman energy as dominant, and at high fields the spectral decomposition is not needed, since the spherical tensor operators are already eigen-operators. The second assumes sufficient symmetry in the bath statistics such that only $J_{l-l}(\omega^l_s)$ are nonzero, and then Hubbard's equation simplifies to

$$R(\sigma) = \sum_{ls} e^{\frac{\beta \omega^l_s}{2}} J_{l-l}(\omega^l_s) \, \text{sech}\big(\beta \omega^l_s / 2\big) \big\{ \big[ (V^l_s)^\dagger \sigma, V^l_s \big] - \big[ \sigma V^l_s, (V^l_s)^\dagger \big] \big\} \qquad (31)$$

In many situations, both conditions are applicable. The double sum in Eq.(29) is useful in zero field. In either case, we have fully reduced Hubbard to Lindbladian form and essentially reproduced the main result obtained by Bengs and Levitt by way of the Lindbladian formalism. One small difference is Hubbard's use of the thermal symmetrizing factor given by the hyperbolic secant function. If so desired, this can be removed as illustrated in Hubbard's paper. It is also imperative to emphasize the importance of the secular approximation in obtaining the Lindblad form. Without this step, relaxation in the rotating frame will be time dependent, with very different, and perhaps even unphysical dynamics. This same approximation is required in a Lindbladian approach, which is often referred to as the "rotating frame" approximation [3].

While very compact, the presence of a finite temperature steady state is definitely obscure. From Eq.(29) directly, the only apparent recourse is to expand the dynamics in a complete set of basis matrices and search for one or more zero eigenvalues. Such a procedure is illustrated in an example with a simple two dimensional density matrix dynamics in [3] where eigenvalues are easily computed.

This is in contrast to Hubbard's original expression, Eq.(14), where the steady state is clearly recognizable, even for arbitrarily large dimensions. Alternatively, we can invoke the secular approximation directly to Eq.(14) and we can enjoy a compromise where one retains the clear presence of the steady state. Hubbard teaches us how to retain explicit information on the fixed point density matrix. However, this seems to be possible only when dynamic frequency shifts can be ignored.

One should also appreciate that a homogeneous system which possess a zero eigenvalue is closely related to an inhomogeneous system. The procedure of homogenizing an inhomogeneous system by incorporating the inhomogeneous vector into equations with an additional dimension, which is invariant with an eigenvalue of zero, has been employed in a Bloch equation analysis of spin echoes [18] , steady state precession [19] and relaxation [20] . Going in the opposite direction can be considerably more difficult.

### 4. **Bloch and Hubbard and Rotating Frame Relaxation:**

We now take a side turn to another aspect of the pioneering work of Bloch and Hubbard, which also has largely gone unrecognized in the NMR literature. As explained in the introduction, the author was recently reacquainted with these aspects in an effort to go beyond a Bloch equation picture with only $R_1$ and $R_2$ for spin locks and adiabatic sweeps in the presence of exchange [21]. Both examples illustrate the use of the high temperature, weak ordering situation that occurs when the full theory contained in Eq.(29) is reduced to the appropriate limit for those circumstances. These applications do not require the Lindbladian form. Nevertheless, it strikes the author as a wasted opportunity to not mention the treatment of rotating frame relaxation by Bloch and Hubbard and therefore reintroduce these two results to 21st century NMR scientists.

At the end of Bloch's paper, he applies his theory to relaxation in the presence of an RF field. For a rank one tensor interaction, such as the fluctuating field relaxation mechanism, he derives a set of generalized Bloch equations in the *rotating frame:*

$$\begin{pmatrix} \dot{M}_x \\ \dot{M}_y \\ \dot{M}_z \end{pmatrix} + \begin{pmatrix} A_x & -\Omega\cos(\theta) & a_x \\ \Omega\sin(\theta) & A_y & -\Omega\sin(\theta) \\ a_z & \Omega\sin(\theta) & A_z \end{pmatrix} \begin{pmatrix} M_x \\ M_y \\ M_z \end{pmatrix} = \begin{pmatrix} c_x \\ 0 \\ c_z \end{pmatrix} \quad (32)$$

When the Rabi frequency, $\Omega\sin(\theta)$ is much smaller than the Larmor frequency $\omega_0$, but still comparable to the resonance offset, $\Omega\cos(\theta)$, we have $a_z = c_x = 0$. At high temperature, $c_z = R_1 M_0$ as usual. In terms of the spectral densities, $J_n(\omega)$ the relaxation parameters are given by the equations

$$A_x = A_y = J_1(\omega_0) + J_0(\Omega) + \big(J_0(0) - J_0(\Omega)\big)cos^2(\theta) \qquad (33)$$

$$A_z = 2J_1(\omega_0) \qquad (33)$$

$$a_x = -\big(J_0(0) - J_0(\Omega)\big)\sin(\theta)\cos(\theta) \qquad (34)$$

In the absence of an RF field, $\theta=0$ and $A_x$ and $A_z$ are $R_2$ and $R_1$ respectively. Note that $a_x$ is generally not zero if the locking field is off resonance.

It is not always appreciated that the usual formulas for rotating frame relaxation are those for the case when the locking field, whose magnitude is given by $\Omega$, is much larger that the relaxation rates. In that situation, a first order perturbation is applicable [21]. If the transformation, denoted by V, diagonalizes the Bloch equations without relaxation, the first order contribution from the relaxation matrix elements is given by the diagonal elements of $V^{-1}RV$, which are then rotating frame relaxation rate constants:

$$\rho_1 = R_1 cos^2(\theta) + \left(R_2 - \left(J_0(0) - J_0(\Omega)\right)\right)sin^2(\theta) \qquad (35)$$

$$\rho_2 = \tfrac{1}{2}R_2 + \tfrac{1}{2}\left(R_2 cos^2(\theta) + (R_1 + J_0(\Omega) - J_0(0))sin^2(\theta)\right) \qquad (36)$$

When the low frequency terms are collected, one obtains expressions that reproduce those for chemical shift exchange, as usually derived from an analysis of exchange perturbation in the limit of fast exchange using the Bloch-McConnell equations [21, 22]. This result is often attributed to Wennerstrom [23]. The theory was already presented in Bloch's paper in 1957. Uncorrelated local fields for two spin ½ systems is an important mechanism for spin isomer conversion, as discussed in Bengs and Levitt.

Bloch presents other applications to his formalism that are of interest. These applications present an excellent catalytic motivation for going through many of his notational details.

After his own exposition and refinement of Bloch's theory, Hubbard also gives an application to rotating frame relaxation, and ups the ante, by considering second rank, dipole-dipole relaxation mechanisms. Hubbard obtains equations for the magnetization dynamics similar to Eq.[32] , with off diagonal contributions. After taking the first order contribution, the rotating frame, spin lattice rate constant is given by

$$\rho_1 = R_1 cos^2(\theta) + R_2 sin^2(\theta) - 6sin^2(\theta)\{-J_0(0) + cos^2(\theta)J_0(\Omega) + sin^2(\theta)J_0(2\Omega)\} \qquad (37)$$

In terms of the spectral densities, $R_1$ and $R_2$ are

$$R_1 = 4\left(J_1(\omega_0) + 4J_2(2\omega_0)\right) \qquad (38)$$

$$R_2 = 6J_0(0) + 10J_1(\omega_0) + 4J_2(2\omega_0) \qquad (39)$$

This result was produced in very different notation by Blicharski in 1972 [24]. Unfortunately, in [24] the various contributions are gathered together in such a manner as to obscure the origin of, and the relationship to each. The reader should keep in mind that for both examples, details regarding scale factors of the spectral densities have been suppressed. These can be added according to the specific needs of their application, be it dipolar, quadrupolar, fluctuating field, or chemical shift exchange.

## 5. Comments and Conclusions:

Given the maturity of the topic of relaxation in magnetic resonance, it is not often that a surprise is forthcoming. Many modern treatments, that are very application oriented, reflect this maturity. For example, the extensive overview offered by Kowalewski and Maler [25] details many of the modern applications. Bloch is not listed in the index, and Hubbard is indexed only in the context of work he did on rotational diffusion applications and the calculation of correlation functions, even though reference [4] is cited in the chapter titled *Redfield Relaxation Theory*. In Redfield's later effort, which appeared as a chapter in *Advances in Magnetic Resonance* [26], Redfield acknowledges the influence of Bloch, and offers his own equations that account for relaxation at finite temperatures, while only citing Hubbard's review article in passing. Redfield makes no effort to demonstrate or expound on the relationship between his expression, and Bloch's or Hubbard's. A glance at Redfield's equation 3.15 in [26] induces one to wonder in what way his result is also Lindbladian, for the expression is very different than Hubbard's equation Eq.(100). The factoring of the spectral densities that Hubbard achieves does not rely on the secular approximation. Nonetheless, a careful study reveals that Redfield's equation is also a mixed Lindbladian type, similar to Eq.(21-22). Here one can fully appreciate the power of using spectral decomposition to factor out the spectral densities, and in doing so, produce an expansion in non-Hermitian operators. Redfield's 1965 result, which *is based on a Hermitian operator expansion* and looks nothing like a Lindbladian, is nonetheless as serviceable as Hubbard's.

Approaches to NMR relaxation theory have changed over its history. In the work of Bloch, Redfield and Hubbard, extensive manipulations are carried out at the level of second order perturbation theory for the *solutions* to the interaction representation density matrix. After extensive manipulations, a finite time step expression is produced, which is argued to be basically the solution to a given differential equation. However, already in the same year that Hubbard's review article appeared in print, Abragam took the alternative approach by directly iterating the differential equation in his treatment of relaxation [27]. This is now the usual practice, and is the path taken, for example, by Goldman in his review of NMR relaxation theory [28], who offers his own treatment of a finite temperature relaxation theory therein and also uses spectral decomposition for that case. This same iteration approach is also adopted by recent expositions using the Lindbladian formalism and weak collision, Markovian bath dynamics and the secular approximation. Again, a good illustration of this is given in [3]. As mentioned earlier, a study of that overview reveals that many of the same tools, e.g. use of Hamiltonian projection operators to obtain eigen-matrices, as used by Bloch and Hubbard, are also brought to bear in the same manner. The historical overview mentioned in section 2 [9], also outlines other open quantum system efforts made by various researchers, and there is a strong enough similarity to suspect that these have rediscovered the main results of Bloch and Hubbard, as well as having anticipated the Lindbladian form.

It is possible to make the argument that NMR theory needs to modernize, in keeping with new approaches that appear to have a more firm foundation in quantum theory. The surprise, is that these new methods can find their own perfect reflection in the best work of the old masters.

Appendix: Some Elaborations on Lindblad and Section 2.

In this appendix, a *sketch* of further aspects of the Lindbladian form is offered. It is self-contained and does not require further references than those already provided.

The most general transformation for a matrix, in particular the density matrix can be written in the form

$$\rho' = \sum_{\alpha\beta\alpha'\beta'} C_{\alpha\beta\alpha'\beta'} \, X^{\alpha\beta} \rho X^{\alpha'\beta'} \qquad (A1)$$

Where the $X^{\alpha\beta}$ are the fundamental basis matrices defined in Section 3. Rather than derive this equation, one can grasp that it is correct by reducing it to component form

$$\rho_{ij}' = \sum_{\beta\alpha'} C_{i\beta\alpha'j} \, \rho_{\beta\alpha'} \qquad (A2)$$

Here, one recognizes the Redfield notation but with slightly rearranged indices. Introducing a complete set of Hermitian matrices $O_k$ provides for a more compact notation where now

$$\rho' = \sum_{kk'} G_{kk'}\, O_k \rho O_{k'} \qquad (A3)$$

If the transformation preserves the Hermitian character of the density matrix, the G is Hermitian in the k indices $G_{kk'} = G_{k'k}{}^*$. If the trace is also invariant then

$$\sum_{kk'} G_{kk'}\, O_{k'}\, O_k = \mathbf{1} \qquad (A4)$$

This is now all we need to obtain a difference equation in $\rho' - \rho$ and obtain the Lindblad form. However this is not usually the way the problem is approached. Instead the starting point is from the "Krauss form", which is obtained by assuming that the matrix G is positive. Being positive, one can factor G in the same manner illustrated in Section 3 via the spectral decomposition for H. This then allows summations over the k indices to produce the Krauss operators for the transformation [3,6]. The one remaining index is over the eigenvalue index. However this is basically going a step to far, and it is more direct to use Eq.(A3). The required subtraction can be implemented by substituting Eq.(A4) for the identity in a symmetrical manner:

$$\rho' - \rho = \sum_{kk'} G_{kk'}\, O_k \rho O_{k'} - \tfrac{1}{2}(\rho\mathbf{1} + \mathbf{1}\rho) \qquad (A5)$$

To complete the process, one now extracts those terms in the operator expansion that involves the identity operator, which we ascribe to the zero index $O_0 = \mathbf{1}$. It is a simple matter to see that these can be collected into the expression

$$\sum_{k} \tfrac{1}{2i}\,(G_{k0} - G_{k0}{}^*)\, i[O_k, \rho] \qquad (A6)$$

This represents the part of the transformation generated by a commutator. The remaining part, where the sums now exclude the identity matrix, is now of the Lindbladian form

$$\rho' - \rho = \sum_{kk'} G_{kk'} \left[ O_k \rho O_{k'} - \tfrac{1}{2}\{\rho,\ O_{k'} O_k\} \right] \qquad (A7)$$

One can then use the positivity of G to factor this expression into non-Hermitian operators. In this way one can see that the approaches from the generalized differential equation and the one based on the general solution are equivalent.

**Acknowledgments:** The author thanks C. Bengs and M. Levitt for friendly correspondence and in particular C. Bengs for pointing out the work of Chruscinski et. al. on the physics archives and for offering some useful suggestions for improving Section 2.

**References:**

1. Christian Bengs and Malcolm Levitt, A master equation for spin systems far from equilibrium, J. Magn. Reson. 310, (2020) 106645.

2. A.G. Redfield, IBM J. Res. Dev.1 (1957) 19-31

3. Daniel Manzano, AIP Advances 10.025106 (2020).

4. F. Bloch, Phys. Rev. 102 (1957) 104-135

5. P. S. Hubbard, Rev. Mod. Phys. 33 (1961) 249-264

6. J.A. Gyamfi, Eur. J. Phys.  (2020)

7. V. Gorini, A. Kossakowski and E.C.G. Sudarshan, J. Math. Phys. 17, (1976) 821-825

8. G. Lindblad, Commun. Math. Phys. 48 (1976) 119-130.

9. D. Chruscinski and S. Pascazio, Physics arXiv: 1710.0593v2.

10. P. R. Halmos, "Finite Dimensional Vector Spaces", 2$^{nd}$ Ed. Dover Books.

11. L.D. Landau and E.M. Liftshitz, Statistical Physics, 2$^{nd}$ Ed. 1977, Pergamon Press (New York)

12. WH. Press, S.A. Teukolsky, W.T. Vetterling and B.P. Flannery, Numerical Recipes in C, Cambridge University Press 1992.

13. P. Pechukas, Phys. Rev. Lett. 73 (1994) 1060

14. A. Shaji and E.C.G. Sudarshan, Phys. Lett. A 341 (2005) 48-56

15. L. Weberlow and R.E. London, Concepts in Mag. Res. 5 (1996) 325-338.

16. S. Alexander, J. Chem. Phys. 37 (1962) 967-974.

17. S. Alexander, J. Chem. Phys. 37 (1962) 974-980.

18. A.D. Bain, C. K. Anand and Z. Nie, J.Magn. Reson. 209 (2011) 183-194

19. I. Nazarova and M.A. Hemminga, J. Magn. Reson. 170 (2004) 284-289.

20. M.H. Levitt and L. Di Bari, Bull. Magn. Reson. 69, (1992) 3124.

21. T.M. Barbara, J. Magn. Reson. 265 (2016) 45-51.

22. D. Abergel and A. Palmer, Concepts in Magnetic Resonance A, 19A (2003) 134-148.

23. H. Wennerstom, Mol. Phys. 24 (1972) 69-80.

24. J.S. Blicharski, Acta Physiol Polo A 41 (1972) 223-236

25. J. Kowalewski and L. Maler, "Nuclear Spin Relaxation in Liquids: Theory, Experiments and

Applications", Taylor and Francis New York, 2006

26. A. G. Redfield, Advances in Magnetic Resonance, Vol. 1, Academic Press, New York (1965)

27. A. Abragam, Principles of Nuclear Magnetism, Oxford University Press, 1983.

28. M. Goldman, J. Magn. Reson. 149 (2001) 325-338